\documentclass[11pt,a4paper]{article}
\usepackage[utf8]{inputenc}
\usepackage[T1]{fontenc}

\usepackage[margin=2cm,a4paper]{geometry}
\usepackage{parskip}

\usepackage{graphicx}
\usepackage{tikz}
\usepackage{epsfig}
\usepackage{wrapfig}
\usepackage{caption}
\usepackage{colortbl}
\usepackage{multirow}

\usetikzlibrary{matrix,shapes}
\usepackage{mathtools}
\usepackage{amsmath}
\usepackage{amssymb}
\usepackage{siunitx}
\usepackage{microtype}
\usepackage{authblk}

\usepackage[colorlinks=true, allcolors=blue]{hyperref}
\usepackage{xr}
\usepackage[style=phys]{biblatex}
\def\equationautorefname#1#2\null{Eq.#1(#2\null)}
\renewcommand{\figureautorefname}{Fig.}

\newcommand{\figref}[2]{\hyperref[#1]{\figureautorefname~\ref*{#1}#2}}

\renewcommand{\Im}{\operatorname{Im}}

\newcommand{\Jmod}{\mathbf{J}^{\mathrm{mod}}}

\title{Incorporating QM/MM molecular dynamics into the few-mode quantization approach for light-matter interactions in nanophotonic structures}

\author[1,2]{Ruth H. Tichauer\thanks{ruth.tichauer@uam.es}}
\author[1,2]{Maksim Lednev}
\author[3]{Gerrit Groenhof}
\author[1,2]{Johannes Feist\thanks{johannes.feist@uam.es}}

\affil[1]{Departamento de Física Teórica de la Materia Condensada, Universidad Autónoma de Madrid, E-28049 Madrid, Spain}
\affil[2]{Condensed Matter Physics Center (IFIMAC), Universidad Autónoma de Madrid, E-28049 Madrid, Spain}
\affil[3]{Nanoscience Center and Department of Chemistry, University of Jyväskylä, P.O. Box 35, Jyväskylä, 40014, Finland}

\begin{document}

\maketitle
\begin{abstract}
In the context of light-matter interactions between organic chromophores and confined photons of (plasmonic) nano-resonators, we introduce a general framework that couples ab initio QM/MM molecular dynamics with few-mode field quantization to simulate light-matter interactions of molecular emitters at the nanoscale.
Arbitrary, lossy, and spatially inhomogeneous photonic environments are represented by a minimal set of interacting modes fitted to their spectral density, while geometry-dependent molecular properties are computed on the fly.
Applications to few-molecule strong coupling show that strong coupling persists when molecular degrees of freedom and disorder are included for the chosen system consisting of a nanoparticle dimer coupled to multiple emitters.
At the same time, symmetry-protected degeneracies of two-level models are lifted.
The framework further reveals how spatial field inhomogeneity and molecular disorder shape cavity-mediated energy transfer, illustrated for an HBQ-Methylene Blue donor-acceptor combination in a five-emitter system.
\end{abstract}

\section{Introduction}
Strong coupling between light and matter has emerged as a powerful framework to control molecular properties and chemical reactivity~\cite{Torma2015,Garcia-Vidal2021}.
When organic molecules are placed inside optical cavities and interact strongly with confined electromagnetic modes, hybrid light-matter states called polaritons form~\cite{Tavis1969}.
These polaritons exhibit unique properties that blend the characteristics of both photons and molecular excitations, opening new avenues for manipulating matter through vacuum fields~\cite{Garcia-Vidal2021}.
While strong coupling was initially demonstrated in Fabry-Pérot microcavities with large ensembles of molecules~\cite{Michetti2005,Agranovich2007}, recent advances in plasmonic nanocavities have enabled strong coupling with only a few~\cite{Heintz2021} or even single molecules~\cite{Chikkaraddy2016} at room temperature.

Despite these experimental advances, theoretical descriptions of light-matter interactions at the nanoscale involving organic emitters face significant challenges due to the inherent complexity of such hybrid systems~\cite{Fregoni2022}.
On the one hand, photoactive molecules possess many degrees of freedom, making their photo-excitation dynamics dependent on both electrons and nuclei through non-adiabatic effects~\cite{CrespoOtero2018,Ehrenfest1927,Tully1990,Granucci2001}.
The coupling of electronic and nuclear motion leads to homogeneous broadening of electronic transitions, while interactions with the surrounding environment induce inhomogeneous broadening.
On the other hand, electromagnetic fields confined at the nanoscale by nano-resonators are highly inhomogeneous, with field amplitudes that depend on both frequency $\omega$ and position $\mathbf{r}$~\cite{Lednev2025,Novotny2012}.
In plasmonic nanocavities, this spatial inhomogeneity is particularly pronounced, with field enhancements concentrated in nanometer-scale ``hot spots''~\cite{Baumberg2019,Bedingfield2023}.

Several computational approaches have been developed in recent years to describe light-matter interactions involving photoactive molecules with high chemical accuracy~\cite{Luk2017,Vendrell2018,Ulusoy2019,Fabri2020,Gudem2021,Fregoni2021}.
These methods range from studies of single molecules in gas phase coupled to single-mode Fabry-Pérot resonators~\cite{Vendrell2018,Ulusoy2019,Fabri2020,Gudem2021} to treatments of molecules near plasmonic nanostructures~\cite{Fregoni2021}.
However, most approaches either neglect environmental effects by considering isolated molecules~\cite{Vendrell2018,Ulusoy2019,Fabri2020,Gudem2021,Fregoni2021} or approximate the electromagnetic environment as a single cavity mode~\cite{Luk2017,Vendrell2018,Ulusoy2019,Fabri2020,Gudem2021}.

To address the complexity associated to material systems, Luk et al.~\cite{Luk2017} pioneered a multiscale approach using Quantum Mechanics/Molecular Mechanics (QM/MM)~\cite{Warshel1976} Molecular Dynamics (MD)~\cite{Frenkel2001} simulations to capture homogeneous and inhomogeneous broadening of molecular transitions as well as chemical reactivity.
In this framework, atoms in the quantum region are treated with \emph{ab initio} electronic structure methods, while the surrounding environment is described with classical force fields.
The time evolution of the hybrid light-matter system follows a mixed quantum-classical dynamics scheme~\cite{CrespoOtero2018,Ehrenfest1927,Tully1990,Granucci2001}, separating fast (electronic and photonic) and slow (nuclear) degrees of freedom following the Born-Oppenheimer-like partitioning scheme leading to polaritonic potential energy surfaces~\cite{Galego2015,Feist2018}.
In this approach, the time-dependent Schrödinger equation is solved for the coupled electron-photon wavefunction $\Psi(t)$ that parametrically depends on the nuclear coordinates $\mathbf{R}$, while nuclear motion is governed by classical equations of motion in the potential landscape created by the fast (electron+photon) subsystem.
Despite its sophistication in treating the molecular subsystem, this original implementation was limited to collective coupling of many molecules to a single cavity mode.

We recently extended this QM/MM MD framework to multiple cavity modes~\cite{Tichauer2021}, enabling the description of the full parabolic dispersion of Fabry-Pérot resonators.
However, single-molecule and few-molecule strong coupling, which has been experimentally achieved in plasmonic nanocavities~\cite{Chikkaraddy2016,Heintz2021}, requires a treatment of the highly inhomogeneous and multimodal electromagnetic field distributions characteristic of such systems~\cite{Baumberg2019,Bedingfield2023}.

In this work, we develop a general computational framework that combines QM/MM MD simulations with the few-mode field quantization approach~\cite{Medina2021,SanchezBarquilla2022} to accurately describe arbitrary nanophotonic structures.
This method quantizes the continuous electromagnetic environment using a minimal set of interacting modes whose parameters are determined by fitting the spectral density obtained from classical electromagnetic simulations.
We apply this framework to investigate how molecular degrees of freedom affect the excitation dynamics of a collection of photo-active molecules in close proximity to plasmonic nanocavities.
Our results show that strong light-matter coupling persists despite dynamic disorder introduced by nuclear motion and environmental fluctuations.
However, symmetry-based degeneracies that are preserved in idealized two-level system (TLS) models are lifted when realistic molecular properties are included.
Furthermore, our approach enables the study of cavity-mediated intermolecular energy transfer with full spatial resolution.

The paper is divided as follows: in \autoref{sec:model} we present the model of light-matter interactions employing just a few modes for the confined electromagnetic field~\cite{Medina2021,SanchezBarquilla2022}, in \autoref{sec:methods} we describe the methods and the systems employed in our simulations and, in \autoref{sec:results}, we present and discuss the obtained numerical results.
We provide a summary and the implications of this work in the conclusions.

\section{Theory}
\label{sec:model}
The Tavis-Cummings (TC) Hamiltonian~\cite{Tavis1969} is a standard starting point to describe an ensemble of $N$ identical two-level emitters coupled to a single cavity mode within the rotating-wave approximation (RWA):
\begin{equation}
\hat H^{TC}=\hbar\omega_e\sum_{i=1}^N\hat{\sigma}_i^\dagger\hat{\sigma}_i + \hbar\omega_c\hat{a}^\dagger \hat{a} - \hbar g\sum_{i=1}^N (\hat{a}^\dagger \hat{\sigma}_i+\hat{a} \, \hat{\sigma}_i^\dagger)\,,
\label{eq:HTC}
\end{equation}
with uniform emitter transition frequency $\omega_e$, single cavity frequency $\omega_c$, and identical light-matter coupling strength $g$ for all emitters.

For organic chromophores placed near nanophotonic structures, almost all the assumptions inherent to the TC model are typically violated: (i) Molecular emitters are not identical two-level systems, since nuclear motion and heterogeneous environments induce time-dependent, molecule-specific excitation energies and transition dipoles. (ii) The electromagnetic environment is strongly multimodal and lossy, making a single-mode description or even a description based on a few independent modes insufficient. (iii) For each photonic mode, the field is highly inhomogeneous, so couplings differ from emitter to emitter.

Instead of undoing the approximations that went into the TC model step by step, we therefore start directly from a much more general description based on few-mode quantization for arbitrary nanophotonic structures~\cite{Feist2021,Medina2021,SanchezBarquilla2022}, which does not rely on any assumptions on the structure or nature of the photonic modes or the emitters, but instead is based on the theory of macroscopic QED~\cite{Scheel2008} to describe arbitrary electromagnetic environments.
The method relies on the observation that the cavity-mediated dynamics and interaction between emitters is fully determined by the spectral density of the electromagnetic environment.
This in turn is determined by the dyadic Green's function $\mathbf{G}(\mathbf{r}_i,\mathbf{r}_j,\omega)$, specifically its imaginary part for reciprocal media, as we will assume here (for non-reciprocal media, it generalizes to the anti-Hermitian part~\cite{Buhmann2012,SanchezMartinez2024General}):
\begin{equation}
\mathbf{J}_{ij}(\omega)= \frac{\omega^2}{\hbar\pi\epsilon_0c^2} \, \Im\big[\mathbf{G}(\mathbf{r}_i,\mathbf{r}_j,\omega)\big].
\label{eq:Jij}
\end{equation}
Here, $\mathbf{r}_i$ and $\mathbf{r}_j$ are the positions of emitters $i$ and $j$, respectively, $\epsilon_0$ is the vacuum permittivity, and $c$ is the speed of light in vacuum.
$\mathbf{J}_{ij}(\omega)$ is a matrix in Cartesian components of the electric field vector for each combination of emitter indices $i$, $j$.
Note that we here chose a slightly different presentation of the method than in previous works, which allows the description of arbitrary dipole orientations more naturally.
We then construct a minimal, interacting few-mode model that reproduces $\mathbf{J}_{ij}(\omega)$ as closely as possible, described by the Hamiltonian
\begin{equation}
H = \sum_{i=1}^N H^\mathrm{e}_{i} + \sum_{k,l} \hbar\omega_{kl} \hat{a}^\dagger_{k} \hat{a}_{l} - \sum_{i,k} \hat{\boldsymbol{\mu}}_i \cdot \mathbf{g}_{ik} (\hat{a}_k + \hat{a}_k^\dagger),\label{eq:H_FM}
\end{equation}
where $H^\mathrm{e}_{i}$ is the Hamiltonian of emitter $i$, $\hat{\boldsymbol{\mu}}_i$ its transition dipole operator, $\hat{a}_k^\dagger$ ($\hat{a}_k$) the creation (annihilation) operator of the quantized photonic mode $k$, $\omega_{kl}$ the mode-mode coupling matrix, and $\mathbf{g}_{ik}$ the electric field strength of mode $k$ at the position of emitter $i$.
The photonic modes are lossy and the system dynamics are then described by a Lindblad master equation,
\begin{equation}
  \dot{\rho} = -i\left[H, \rho\right] + \sum_k \kappa_k L_{a_k}[\rho],
\end{equation}
where $\rho$ is the system density matrix, $L_O[\rho] = O \rho O^{\dagger} - \frac12\{O^{\dagger}O,\rho\}$ is a Lindblad dissipator, and $\kappa_k$ is the decay rate of mode $k$. The spectral density of this few-mode model is given by~\cite{SanchezBarquilla2022}
\begin{equation}
		\Jmod_{ij}(\omega)=\frac{1}{\pi}\sum_{kl} \mathbf{g}_{ik} \Im\!\left[\frac{1}{\mathbf{\tilde H}-\omega}\right]_{kl} \mathbf{g}_{jl},
		\label{eq:Jmod}
\end{equation}
where $\mathbf{\tilde H}_{kl} = \omega_{kl} - \frac{i}{2} \kappa_k\delta_{kl}$.
The real parameters $\omega_{kl}$, $\kappa_k$, and $\mathbf{g}_{ik}$ of the few-mode quantization model are then obtained by nonlinear fitting of the effective spectral density to the full one from \autoref{eq:Jij}, calculated by solving Maxwell's equations.
This procedure yields a compact photonic Hamiltonian without assuming spatial homogeneity, a single mode, or identical emitters.

Up to this point, the nature of the emitter has not been specified---it is simply described by its Hamiltonian and dipole operator.
In most previous works on few-mode quantization~\cite{Feist2021,Medina2021,SanchezBarquilla2022,Lednev2024,Lednev2025,SanchezMartinez2024Mixed}, the emitters were modeled as two-level systems.
Here, we go beyond this simplified description and model the molecular subsystem using the approach of QM/MM molecular dynamics for polaritonic systems~\cite{Luk2017,Tichauer2021}.
In this framework, each molecule $i$ is treated as a few-level system (restricted to two levels in the applications below) whose properties depend on the nuclear coordinates $\mathbf{R}_i$ of that molecule at each time step of the MD simulation.
We thus compute on the fly the geometry-dependent ground- and excited-state energies $V_{S_0}(\mathbf{R}_i)$ and $V_{S_1}(\mathbf{R}_i)$, as well as the transition dipole $\boldsymbol{\mu}_i(\mathbf{R}_i)$ between them.
Applying the RWA to the general light-matter Hamiltonian then leads to our working model for the light-matter dynamics,
\begin{equation}
  \begin{split}
    H^{LM} &= \sum_{i=1}^N \left[V_{S_0}(\mathbf{R}_i) + \hbar\omega^\mathrm{e}_i(\mathbf{R}_i) \hat{\sigma}_i^\dagger\hat{\sigma}_i\right]
    + \sum_{kl} \hbar\omega_{kl} \hat{a}_k^\dagger \hat{a}_l 
    - \sum_{i,k}\boldsymbol{\mu}_i(\mathbf{R}_i)\cdot \mathbf{g}_{ik} \left(\hat{\sigma}_i^- \hat{a}_k^\dagger + \hat{\sigma}_i^+ \hat{a}_k\right),
  \end{split}
\label{eq:HLM}
\end{equation}
where $\hbar\omega^\mathrm{e}_i(\mathbf{R}_i) = V_{S_1}(\mathbf{R}_i)-V_{S_0}(\mathbf{R}_i)$ is the excitation energy of molecule $i$ at geometry $\mathbf{R}_i$, and $\hat{\sigma}_i^+$ ($\hat{\sigma}_i^-$) is the raising (lowering) operator between the electronic ground and excited state of molecule $i$.
In contrast to the TC model, this Hamiltonian retains the multimode structure of the EM field, spatially varying couplings, and the full molecular complexity captured by QM/MM MD\@.

We perform simulations starting with a single excitation in the system. Since the Hamiltonian conserves the number of excitations, and the only terms that do not conserve the number of excitations in the Lindblad master equation are the so-called quantum jump or refilling terms $\propto a_k \rho a_k^\dagger$ (which make population that decays from the single-excitation subspace ``reappear'' in the ground state manifold), the dynamics within the single-excitation subspace is equivalently determined by the non-Hermitian Schrödinger equation with Hamiltonian $\tilde{H}^{LM} = H^{LM} - \frac{i}{2} \sum_k \kappa_k \hat{a}_k^\dagger \hat{a}_k$. We thus work with a wave-function based approach within the single-excitation subspace, expanding the total electron-photon wavefunction as $|\Psi(t)\rangle = \sum_{m=1}^{N+n} c_m(t) |\psi^m\rangle$, where $|\psi^m\rangle$ is the state with a single excitation in component $m$ of the system and all others in their ground state ($m=1,\ldots,N$ being the molecules, and $m=N+1,\ldots,N+n$ being the photonic modes), and $c_m(t)$ is its time-dependent amplitude.

To incorporate the influence of the electromagnetic field on nuclear motion, we augment the molecular dynamics force evaluation with contributions arising from light-matter coupling to the confined modes.
Within the mean-field Ehrenfest framework~\cite{Ehrenfest1927,CrespoOtero2018,Agostini2019}, each atom $a$ in molecule $j$ experiences a Hellmann-Feynman force gradient that supplements the electronic (QM) or force-field (MM) potential:
\begin{equation}\label{eq:fMF}
 \mathbf{F}_{{HF}\to a\in j}^{MF} = |c_j(t)|^2\nabla_{a}V_{{S}_1}^j({\bf{R}}_j)+ \sum_{m\neq j}^{N+n}|c_m(t)|^2\nabla_{a}V_{{S}_0}^j({\bf{R}}_j) 
    - \sum_{k=0}^{n-1}\left(c_j^*(t) c_k(t)\nabla_{a}\boldsymbol{\mu}({\bf{R}}_j)\cdot\mathbf{g}_{jk} + c.c.\right).
\end{equation}
\autoref{eq:fMF} applies when working in a diabatic basis $\{|\psi^m\rangle\}$ of the hybrid light-matter system, constructed as the tensor product of uncoupled molecular states $|S_0^0 \ldots S_1^i \ldots S_0^N\rangle$ and photonic Fock states $|0_0 \ldots 1_k \ldots 0_{n-1}\rangle$~\cite{Tichauer2021,Sokolovskii2024}.
This diabatic basis is equivalent to the unique adiabatic basis consisting of the polaritonic eigenstates of the full light-matter Hamiltonian in \autoref{eq:HLM}.
In section I of the Supplementary Information (SI), we provide the corresponding Hellmann-Feynman force expressions in the adiabatic representation.
Both formulations are implemented in our simulation code.

\section{Methods}
\label{sec:methods}
\subsection{Organic chromophores}
\label{sec:model-mols}
We investigate three distinct photoactive molecules: Rhodamine (Rho), Methylene Blue (MeB), and 10-hydroxybenzo[h]quinoline (HBQ), whose structures are shown in the insets of \autoref{fig:TLS}(a) and \autoref{fig:hbq-meth-gap}(a).
Prior to the few-mode QM/MM MD simulations described in \autoref{sec:results}, all molecular systems underwent a two-stage preparation protocol: initial optimization, solvation, and equilibration at the classical molecular mechanics (MM) level, followed by refinement at the QM/MM level.

For the classical MM simulations, we employed the Amber03 force field~\cite{Duan2003} combined with the TIP3P water model~\cite{Jorgensen1983} for Rhodamine, using parameters developed by Luk et al.~\cite{Luk2017}.
Methylene Blue utilized parameters from Sokolovskii et al.~\cite{Sokolovskii2024b}, while HBQ solvated in cyclohexane was described with the GROMOS-2016H66 force field~\cite{Horta2016}.
Detailed protocols for the MM-level optimization, solvation, and equilibration of Rhodamine can be found in Tichauer et al.~\cite{Tichauer2021}, and those for Methylene Blue and HBQ in Sokolovskii et al.~\cite{Sokolovskii2024b}.

Following MM preparation, all systems were equilibrated at the QM/MM level using an MD time step of 1 fs.
Consistent with previous implementations~\cite{Luk2017,Tichauer2021}, the QM region of Rhodamine comprised the three fused rings, treated at the restricted Hartree-Fock (RHF) level with the 3-21G basis set.
The remainder of the molecule (benzene side group) and the aqueous solvent constituted the MM region, described by the Amber03 force field~\cite{Duan2003} and TIP3P water model~\cite{Jorgensen1983}, respectively.
For Methylene Blue and HBQ, the entire molecular framework was treated quantum mechanically.
Ground-state electronic properties were computed using density functional theory (DFT)~\cite{Hohenberg1964} with the $\omega$B97X-D~\cite{Chai2008} and CAM-B3LYP~\cite{Becke1993,Yanai2004} functionals for Methylene Blue and HBQ, respectively, both employing the 6-31G* basis set~\cite{Ditchfield1971}.
Methylene Blue was solvated in water (TIP3P model~\cite{Jorgensen1983}), while HBQ was solvated in cyclohexane (GROMOS-2016H66 force field~\cite{Horta2016}).
To account for environmental electrostatic effects, the QM regions of Rhodamine and Methylene Blue were electrostatically embedded in their respective MM environments, experiencing the Coulomb field of all MM atoms within cutoff radii of 1.6 nm and 1.0 nm, respectively.
Lennard-Jones interactions between QM and MM atoms provided mechanical coupling for all systems.
Temperature control was achieved using the velocity-rescaling thermostat~\cite{Busi2007} with a coupling time constant of 0.1 ps, applied every 100 MD steps.

After QM/MM system preparation, excited-state potential energy surfaces $V_{S_1}^{QM/MM}(\mathbf{R})$ were constructed on-the-fly during the few-mode QM/MM Ehrenfest MD simulations by computing the first singlet excited state ($S_1$) at each time step.
The QM/MM partitioning schemes established for ground-state calculations were maintained for excited-state treatments.
For Rhodamine, the QM region was described using configuration interaction singles (CIS) with the 3-21G basis set, while Methylene Blue and HBQ employed time-dependent DFT (TD-DFT)~\cite{Runge1984} with the $\omega$B97X-D and CAM-B3LYP functionals, respectively, both with the 6-31G* basis set.
The MM environments were modeled using the Amber03 force field~\cite{Duan2003} with TIP3P water model~\cite{Jorgensen1983} for Rhodamine, TIP3P water model alone for Methylene Blue, and the GROMOS96-2016H66 force field~\cite{Horta2016} for the cyclohexane solvent surrounding HBQ\@.

All molecular simulations, both in vacuum and within the nano-resonator, were performed using GROMACS 4.5.3~\cite{Hess2008} interfaced with TeraChem~\cite{Sisto2014} for the QM/MM treatment of the material subsystem within the hybrid light-matter system.

\subsection{Photonic resonator}
\label{sec:resonator}
The photonic resonator employed in this work consists of a silver nanosphere dimer oriented along the $z$-axis, with each sphere having a radius of 20 nm and the two particles separated by a 1.5 nm gap (see \figref{fig:TLS}{a} and \figref{fig:hbq-meth-gap}{a}).

Using the Drude model for the frequency-dependent relative permittivity $\epsilon(\mathbf{r},\omega)$ of silver, with literature values for the different parameters~\cite{Palik1998,Delga2014}, and a permeability $\mu(\mathbf{r},\omega)=1$, we computed the electromagnetic dyadic Green's function $\mathbf{G}(\mathbf{r}_i,\mathbf{r}_j,\omega)$ for this nanostructure using the SCUFF-EM boundary element method code~\cite{SCUFF1,SCUFF2}.
In our model geometry, five emitters are positioned within the dimer gap: one at the center and four 1 nm away at symmetrically equivalent peripheral locations forming a square centered on the first emitter (\figref{fig:TLS}{a} and \figref{fig:hbq-meth-gap}{a}).
All emitters lie in the equatorial plane of the gap, with a minimum distance of $0.75~$nm from each nanoparticle surface to prevent quantum tunneling that would arise from wavefunction overlap between the molecular emitters and the metallic nanoparticles.

Following numerical computation of $\mathbf{G}(\mathbf{r}_i,\mathbf{r}_j,\omega)$, we fitted the resulting spectral density $\mathbf{J}_{ij}(\omega)$ (\autoref{eq:Jij}) using a 40-mode representation.
Both the exact and fitted spectral density for the central emitter with its dipole moment oriented along the $z$-axis are displayed in \figref{fig:TLS}{b}.
The fitting procedure was iterated until achieving a convergence criterion of $10^{-4}$ in the relative error.

\section{Results \& discussion}
\label{sec:results}

\subsection{Lindblad master equation vs few-mode QM/MM MD for TLSs}
\label{sec:TLS}
\begin{figure}[htb!]
\includegraphics[width=\textwidth]{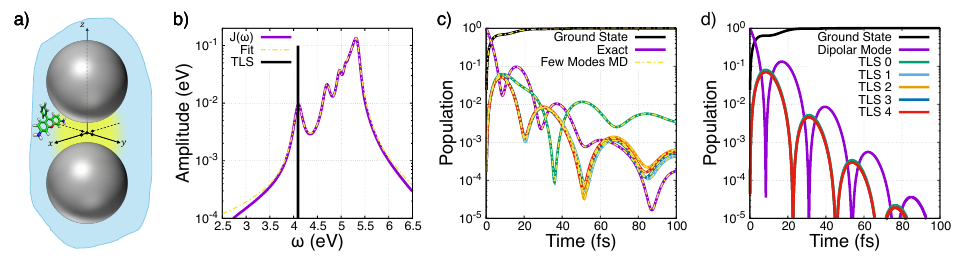}
\caption{(a) Schematic representation of five point-like emitters within the gap of a silver nano-particle dimer oriented along the $z-$axis.
When accounting for the molecular degrees of freedom, a Rhodamine model depicted in the inset is placed at each spot.
(b) Physical (plain purple line) and fitted (yellow pointed-dashed line) spectral densities of the central emitter with its dipole moment oriented along the $z$-axis. The absorption of the emitters corresponds to a delta function (black vertical line) when treated as ideal two-level systems (TLSs).
(c) Time-evolution of the population of the five TLSs resonantly coupled to the dipolar mode of the optical resonator.
(d) Same population dynamics but when the confined electromagnetic field within the nano-cavity is described by a single mode.}
\label{fig:TLS}
\end{figure}

To validate the few-modes QM/MM MD implementation, we start by modeling the five emitters in the setup described in \autoref{sec:resonator} as identical ideal TLSs and compare the dynamics of the excited state populations of this hybrid light-matter system to the dynamics obtained employing the well-established Lindblad master equation of open quantum systems~\cite{Breuer2007}.
These five TLSs are set to have an excitation energy of 4.10 eV and transition dipole moment equal to 9.42 D, comparable to our Rhodamine molecular model.
They are resonantly coupled to the bright dipolar mode~\cite{Delga2014} of the silver nanosphere dimer (see \figref{fig:TLS}{b}).
The confined electromagnetic field within this nano-resonator is modeled with 40 modes such that the first one corresponds to the dipolar mode.
This well separated peak can be fitted with a single Lorentzian at $\hbar\tilde{\omega}_0=4.10$~eV with a width of $\hbar\kappa_0=160$~meV that corresponds to a decay time of $1/\kappa_0 = 8.23$~fs.
We fix the parameters of $\Jmod(\omega)$ (\autoref{eq:Jmod}) such that this fitted dipolar mode is decoupled from the rest of the modes, i.e., the off-diagonal terms in the first row/column of $\mathbf{\tilde H}$ are set equal to zero ($\tilde{\omega}_{0i}=\tilde{\omega}_{i0}=0$ for $i \in [1,n-1]$ where $n=40$ is the number of photonic modes).

Agreement with error below $10^{-4}$ is observed between the few modes-QM/MM MD implementation and the Lindblad master equation in the population dynamics of the five TLSs coupled to the 40-modes silver nanoparticle dimer (see \figref{fig:TLS}{c} and Fig. S1).
While the initial excitation of the bright dipolar mode (purple line in \figref{fig:TLS}{c}) inevitably decays over time, concomitant with a monotonous increase of the ground state occupation (black line) that reflects the total losses in the system, the population of this dipolar mode oscillates coherently with that of the five TLSs, evidencing strong light-matter coupling~\cite{Li2017}.
In other words: as the population of the dipolar mode decreases, that of the TLSs increases, and vice-versa.
The population of the central emitter (green line) differs from that of the four peripheral emitters (light blue, orange, dark blue and red lines), which remain identical within a TLS description.
Their equivalence can be understood by their symmetrically equivalent positions with respect to the central emitter and the silver nanoparticle dimer.
Small differences arise from the non-symmetric meshing of the nanoparticle dimer used to numerically compute the dyadic Green's function $G(\mathbf{r}_i,\mathbf{r}_j,\omega)$.

We additionally computed the excitation dynamics when modeling the confined electromagnetic field with a single mode (\figref{fig:TLS}{d}).
This single mode corresponds to the dipolar mode to which the five emitters are resonantly coupled and has the same parameters as the decoupled fitted dipolar mode.
The population dynamics in \figref{fig:TLS}{d} show that a single-mode approximation is far from accurate for this system.
Even though in the 40-mode fit $\Jmod(\omega)$, the dipolar mode $(\tilde{\omega}_0,\kappa_0)$ is decoupled from the other 39 modes and dominates radiation to the far field~\cite{Delga2014}, a single-mode model cannot reproduce the clear difference in dynamics between the central and peripheral emitters.
The remaining 39 fitting modes ($\{\tilde{\omega}_k,\kappa_k\}$), together with their mutual couplings $\{\tilde{\omega}_{kp}\}$, are responsible for representing the highly inhomogeneous electromagnetic field within the dimer gap that is essential to capture the correct dynamics.
In a single-mode description, the difference between emitters is reduced to a different coupling strength to that mode, which cannot capture the full complexity of the system.
We note that if the coupling to non-resonant modes is weak enough to be perturbative, treating them within a Markovian approximation is a practical alternative that can retain the qualitative physics while enabling an effective single-mode treatment~\cite{SanchezMartinez2024Mixed}.
We do not explore this possibility further here, as our focus is on the full few-mode description. Moreover, since for organic chromophores the excitation energies fluctuate due to nuclear motion and environmental effects, a distinction between resonant and non-resonant modes is not straightforward.

\subsection{Effect of incorporating molecular degrees of freedom}
Having validated the few-mode QM/MM MD approach for TLSs and quantified the accuracy of our implementation against the Lindblad master equation (see \figref{fig:TLS}{c} and Fig. S1), we next examine how molecular degrees of freedom (homogeneous broadening) and solvent interactions (inhomogeneous broadening) affect light-matter interactions.

\begin{figure}[htb!]
\includegraphics[width=\textwidth]{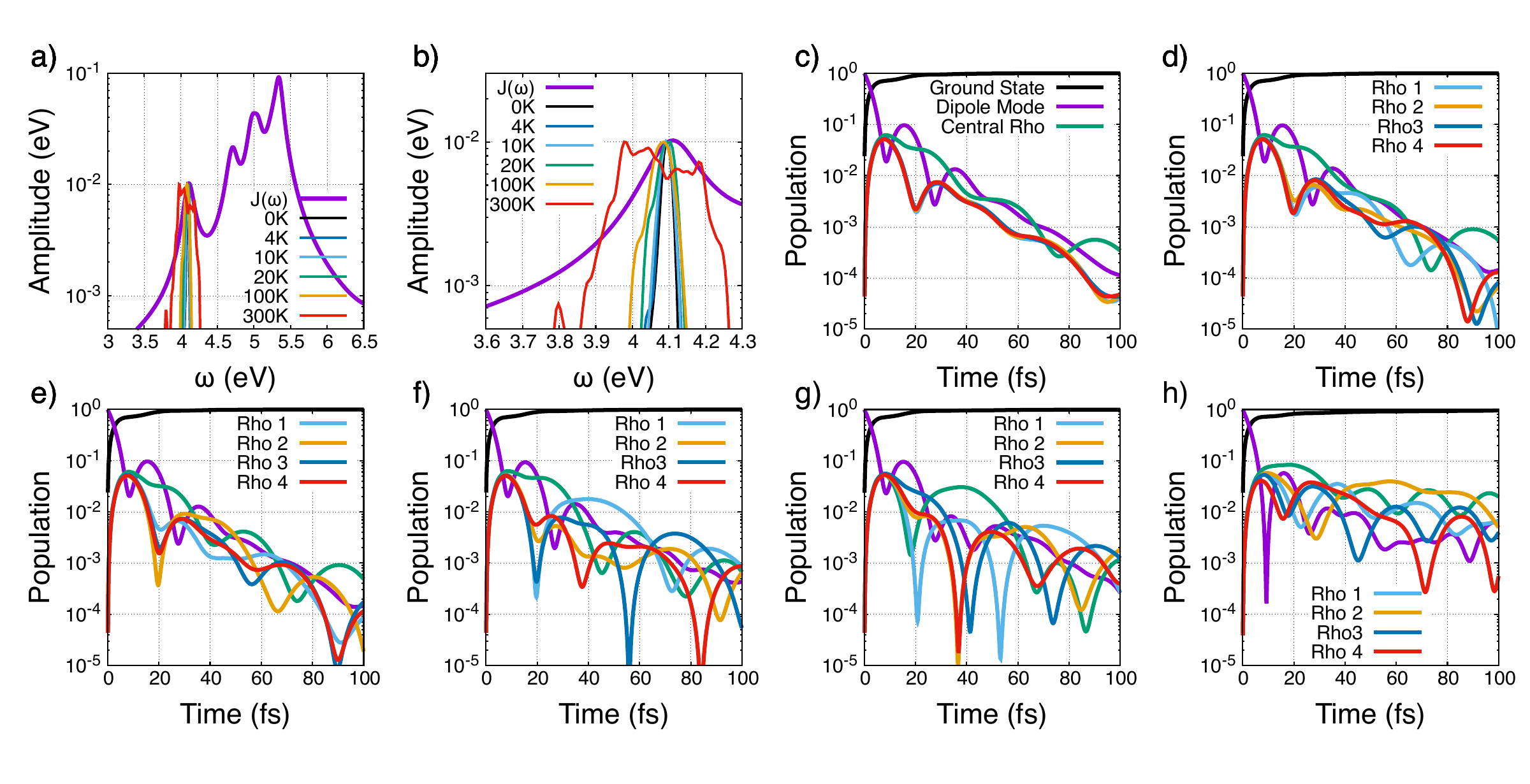}
\caption{(a) Fitted spectral densities of the hybrid light-matter system shown alongside the absorption of the molecular emitters in gas phase at various temperatures and in solution at 300 K.
	(b) Zoom into the dipolar mode of the spectral density and its overlap with the molecular absorption under different conditions.
	(c) Time evolution of the populations of the Rhodamines, the dipolar mode of the silver nanoparticle dimer, and the overall ground state (system losses) for gas-phase emitters at 0 K.
	(d) Same populations at 4 K, (e) 10 K, (f) 20 K, (g) 100 K, and (h) in solution (water, QM/MM model) at 300 K.}
\label{fig:gaz-sol}
\end{figure}
To assess the impact of internal molecular degrees of freedom, we first employ our Rhodamine model in the gas phase (full QM simulations, no solvent) at several temperatures (0, 4, 10, 20, and 100 K).
As described in \autoref{sec:model-mols}, the ground ($S_0$) and first excited ($S_1$) states of the QM region (fused rings of Rhodamine) were treated at the RHF/3-21G and CIS/3-21G levels of theory, respectively.
Although the five emitters share the same initial conformation (i.e., no initial diagonal or off-diagonal disorder as in the TLS case), including internal nuclear motion results in markedly different population dynamics compared to a TLS treatment (\figref{fig:TLS}{c}; \autoref{fig:gaz-sol}).

At 0 K, no vibrational modes are populated initially.
Yet, differences emerge between the TLS treatment and the full-QM gas-phase simulations (\figref{fig:TLS}{c} and \figref{fig:gaz-sol}{c}).
These differences stem from the finite kinetic energy acquired during MD\@.
During the few-mode QM MD simulation, the initial energy deposited in the resonator dipolar mode is partially converted into vibrational energy through light-matter interactions with the confined field.
Because the field enhancement is maximal at the center of the gap, the central Rhodamine attains an average kinetic energy corresponding to an effective temperature of $\sim1.5$~K, while the peripheral molecules heat to $\sim0.25$~K (see Fig. S5).
This temperature difference reflects the unequal electromagnetic field intensities experienced within the nanocavity.
By adjusting the coupling time constant of the velocity-rescaling thermostat, atomic velocities can be rescaled so that the average kinetic energy matches the target temperature.

Increasing the temperature in gas-phase simulations (4, 10, 20, and 100 K; panels (d)-(g) in \autoref{fig:gaz-sol}) progressively lifts the degeneracy among the four peripheral emitters, whose populations cease to oscillate coherently earlier in time ($\sim$35, 30, 15, and 12 fs for 4, 10, 20, and 100 K, respectively).
This effect originates from dynamic disorder.
Different initial atomic velocities, randomly sampled from a Maxwell-Boltzmann distribution, drive distinct explorations of the vast phase space of Rhodamine (37 atoms, yielding 105 vibrational modes).
At 0 and 4 K (panels (c) and (d)), the effective light-matter coupling is reduced relative to the TLS case, as evidenced by less pronounced population oscillations, whereas at 10, 20, and 100 K the coupling is enhanced.
From 10 K onward, the increased homogeneous broadening due to higher vibrational occupation improves the spectral overlap between the molecular excited-state resonance and the resonator dipolar mode (panels (a) and (b)), thereby strengthening the effect of the light-matter coupling such that it overcomes the additional disorder introduced by nuclear motion.
Strong coupling is maintained across all temperatures considered.

\begin{figure}[htb!]
\includegraphics[width=\textwidth]{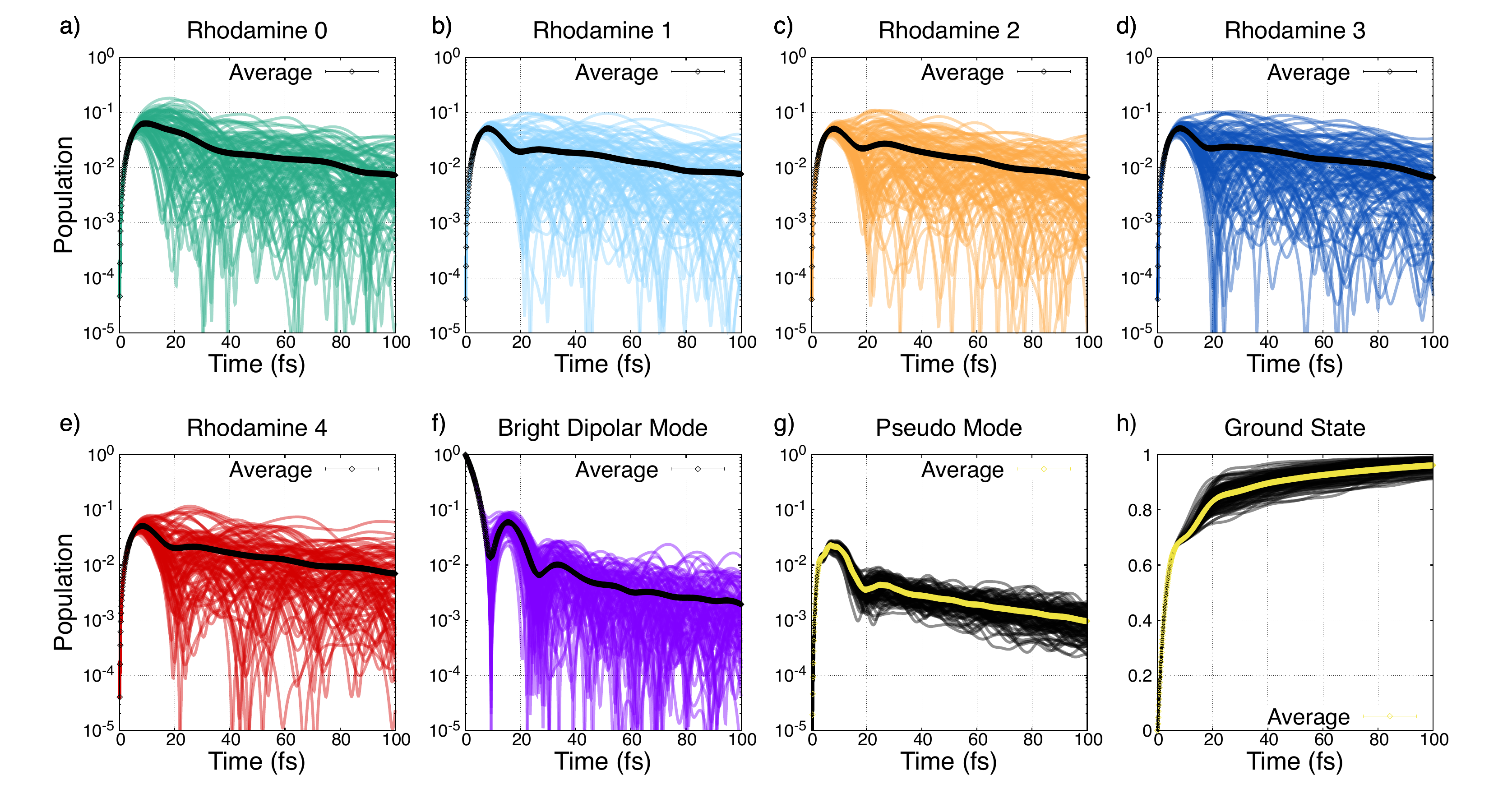}
\caption{(a) Time evolution of the excited-state population of the central Rhodamine molecule within the setup of \figref{fig:TLS}{a} across 101 few-mode QM/MM MD simulations.
The average over these 101 realizations is shown with black diamonds.
	(b), (c), (d), and (e) Time evolution of the populations of the four peripheral molecular emitters.
	(f) Population dynamics of the dipolar mode of the silver nanoparticle dimer.
	(g) Population dynamics of the modes forming the broad pseudo-mode (peak at $\sim$5.35 eV in \figref{fig:TLS}{b}).
The average over the 101 realizations is shown with yellow diamonds.
	(h) Ground-state occupation reflecting losses of the silver nanoparticle dimer.
Note the linear scale in this panel.}
\label{fig:dof}
\end{figure}
Including inhomogeneous broadening by modeling the five emitters in solution (explicit water using the TIP3P model at 300 K) further lifts the degeneracy among the four peripheral molecules, whose excited-state populations remain superimposed only during the first $\sim$7 fs (\figref{fig:gaz-sol}{h}).
Strong coupling persists despite the substantial disorder introduced by the solvent.
However, at 300 K in solution, the interaction strength is reduced compared to gas-phase simulations at 10, 20, and 100 K\@.
The molecular absorption (red curve in \figref{fig:gaz-sol}{b}) broadens sufficiently that the ensemble explores regions of phase space no longer resonant with the resonator dipolar mode.

As mentioned above, because of the vast phase space available to molecular emitters, we conducted 100 additional few-mode QM/MM MD simulations of solvated Rhodamine within the dimer gap.
As shown in \autoref{fig:dof}, coherent population dynamics are maintained in individual realizations but, when averaged, persist only for $\sim$20-30 fs, with two pronounced oscillations for the molecules (panels (a)-(e)) and three for the bright dipolar mode (panel (f)).
In individual realizations the degeneracy among the four peripheral emitters is lifted by distinct nuclear trajectories, but on average the four peripheral positions are equivalent.
Nevertheless, the dynamic disorder introduced by solution-phase MD is insufficient to make the central and peripheral dynamics equivalent.
Thus, despite the small spatial extent of the ensemble (the four peripheral emitters are only 1 nm from the central one), they experience, on average, different electromagnetic environments within the nanoparticle dimer gap.
A single-mode approximation would miss this key feature, which is critical for the design of molecule-nanocavity devices.

\subsection{Energy transfer mediated by a nanocavity}
\label{sec:enertrans}
Organic semiconductor materials are increasingly integrated into electronic technologies, despite inherent disorder that limits efficiency~\cite{Tessler2009}.
For the performance of organic solar cells, LEDs, and transistors, energy transfer between chromophores is often crucial.
In this context, Gonzalez-Ballestero et al.~\cite{GonzalezBallestero2015} predicted that exciton transfer between two-level emitters can be enhanced under strong coupling to nanocavity hot spots.
Leveraging our approach that goes beyond a TLS description of the emitters, we thus also investigate cavity-mediated intermolecular energy transfer.

\begin{figure}[htb!]
\includegraphics[width=\textwidth]{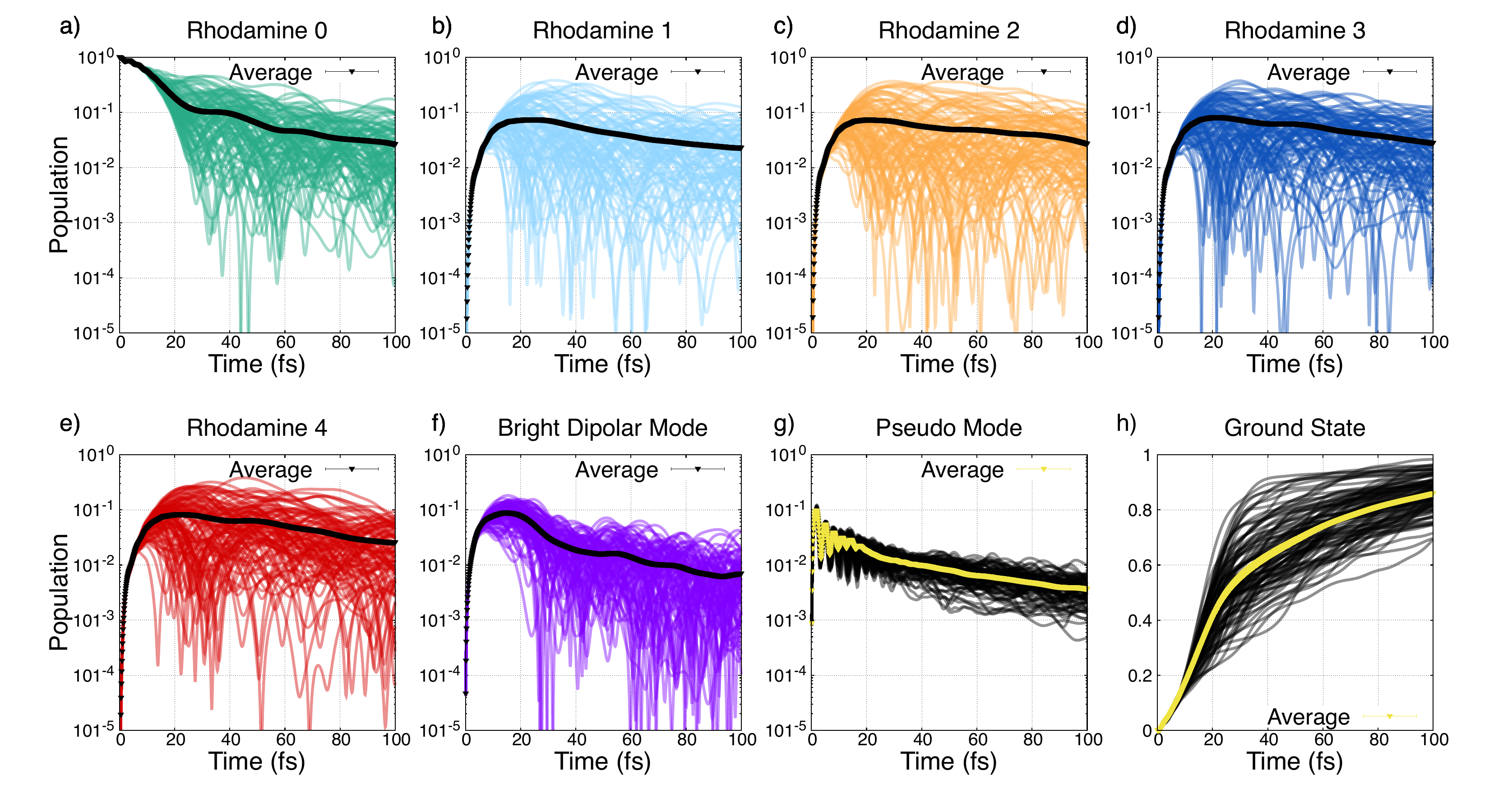}
\caption{(a) Time evolution of the excited-state population of the central Rhodamine in the setup of \figref{fig:TLS}{a} for 100 few-mode QM/MM MD simulations with initial excitation on the central molecule.
The average over these 100 realizations is shown with black diamonds.
	(b), (c), (d), and (e) Time evolution of the populations of the four peripheral molecular emitters.
	(f) Population dynamics of the dipolar mode of the silver nanoparticle dimer.
	(g) Population dynamics of the modes forming the broad pseudo-mode (peak at $\sim$5.35 eV in \figref{fig:TLS}{b}).
The average over the 100 realizations is shown with yellow diamonds.
	(h) Ground-state occupation due to losses of the silver nanoparticle dimer.
Note the linear scale in this panel.}
\label{fig:rho-gap-Etransfer}
\end{figure}

In the setup of \figref{fig:TLS}{a}, rather than initially exciting the bright dipolar mode that couples to the far field~\cite{Delga2014}, we consider an initial excitation localized on the central Rhodamine chromophore.
The populations of the molecular emitters, the resonator dipolar mode, and the overall ground state across 100 few-mode QM/MM MD simulations are shown in \autoref{fig:rho-gap-Etransfer}.

Relative to a TLS treatment (Fig. S6), energy transfer from the central Rhodamine to the peripheral ones is significantly enhanced when molecular degrees of freedom are included.
In the TLS case, peripheral emitters reach at most $\sim$4-5\% occupation by 15 fs and decay to $\sim$2\% by the end of the simulation, whereas with nuclear motion and solvent interactions (\autoref{fig:rho-gap-Etransfer}) individual realizations reach up to $\sim$40\% ($\sim$10\% on average) at $\sim$30-40 fs before decaying to $\sim$10\% ($\sim$3-4\% on average) by 100 fs.
This enhancement arises from increased spectral overlap between Rhodamine absorption and the resonator dipolar mode once homogeneous and inhomogeneous broadening are included.

Since the fluorescence lifetime of Rhodamine is on the order of nanoseconds, radiative molecular decay to free-space background modes is negligible on the timescale of our simulations (hundreds of femtoseconds).
Consequently, both intermolecular energy transfer and overall excitation decay are mediated mostly through the nanocavity modes, and in particular the dipolar mode, $(\tilde{\omega}_0,\kappa_0)$.
The average loss rate is therefore reduced when initially exciting the central Rhodamine instead of the dipolar mode: on average it takes $\sim 80~$fs for the ground state population to reach $80\%$ (\figref{fig:rho-gap-Etransfer}{h}), compared to $\sim 15$~fs when initially exciting the dipolar mode (\figref{fig:dof}{h}).
We note, however, that initially exciting the central Rhodamine (\autoref{fig:rho-gap-Etransfer}) leads to an initial $\sim 10\%$ occupation of the so-called pseudo-mode (\figref{fig:rho-gap-Etransfer}{g}), i.e., the broad peak at $\sim 5.35$~eV that is due to the combined action of high-order modes of the nanoparticles~\cite{Delga2014}, compared to $\sim$2-3\% in \figref{fig:dof}{g}.
Thus, while total losses are reduced, nonradiative channels mediated by the pseudo-mode play a more prominent role for localized initial excitations.

\subsection{Photochemistry-driven energy transfer}
\label{sec:photochem}
\begin{figure}[!htb]
\includegraphics[width=\textwidth]{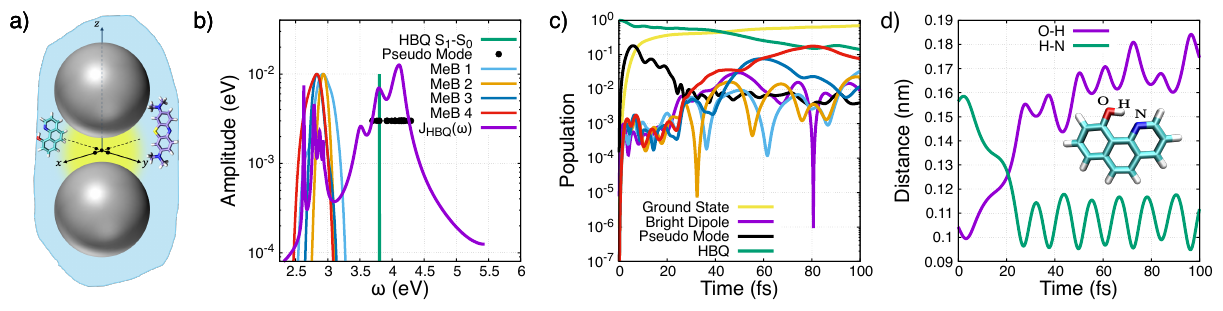}
\caption{(a) Schematic of five molecular emitters within the gap of a silver nanoparticle dimer oriented along the $z$-axis.
HBQ (left inset) occupies the central site, and Methylene Blue molecules (right inset) occupy the peripheral positions.
	(b) Effective spectral density perceived by HBQ shown with the absorption of Methylene Blue (light blue, orange, dark blue, and red lines).
Black dots indicate the fitted-mode frequencies $\tilde{\omega}_k$ forming the pseudo-mode in our model.
	(c) Time evolution of the excited-state populations of the five emitters (HBQ: green; MeB1: light blue; MeB2: orange; MeB3: dark blue; MeB4: red), together with that of the bright dipolar mode (purple) and the pseudo-mode (black; broad peak at $\sim$4 eV).
The ground-state population (yellow) increases steadily due to system losses.
	(d) O-H and H-N distances in HBQ (inset).
An H-N distance of $\sim$0.105 nm corresponds to a bonded state.}
\label{fig:hbq-meth-gap}
\end{figure}

To mimic experimental conditions with substantial disorder at room temperature (300 K), the Rhodamine chromophores in the simulations shown in \autoref{fig:rho-gap-Etransfer} were assigned different initial conformations and velocities, and thus different excitation energies.
Experimentally, however, selectively exciting the central Rhodamine (or a specific peripheral emitter) would be challenging because, on average, the emitters share the same absorption spectrum (Fig. S6b).
We therefore replace the central Rhodamine by HBQ (left inset, \figref{fig:hbq-meth-gap}{a}) and the peripheral emitters by Methylene Blue (right inset, \figref{fig:hbq-meth-gap}{a}).
We also shift the fitted spectral density $\Jmod(\omega)$ associated with the silver nanoparticle dimer so that our Methylene Blue model is resonant with the dipolar mode ($\tilde{\omega}_0,\kappa_0$), here set to $\tilde{\omega}_0=2.89$ eV (Fig. S9). Because we employ the RWA and restrict the excitation dynamics to the single excitation sub-space, we can arbitrarily shift the spectral density by a constant value. 

Upon far-blue/UV absorption, HBQ undergoes an ultrafast excited-state intramolecular proton transfer reaction~\cite{Kim2009,Lee2013}.
The associated conformational change (H bonded to N instead of O; see inset in \figref{fig:hbq-meth-gap}{d}) produces a large Stokes shift with emission peaking in the green-yellow region.
Because the absorption maximum of Methylene Blue is in the same spectral region (Fig. S7), this donor-acceptor pair is well suited to study photochemistry-driven, cavity-mediated intermolecular energy transfer, similar to the recent proposal of photochemistry-driven polariton propagation by Sokolovskii et al.~\cite{Sokolovskii2024b} using the same molecular pair.
Critically, our approach enables photochemical reactions in cavities owing to the QM/MM treatment of the material subsystem.
Whereas prior implementations were restricted to 2D Fabry-Pérot microresonators~\cite{Tichauer2021}, the present extension accommodates photonic resonators of arbitrary shape.

As shown in Fig. S8, a TLS treatment precludes energy transfer due to minimal spectral overlap between donor and acceptor (HBQ absorption at $\sim$3.8 eV, MeB at $\sim$2.8 eV in our model).
While within a TLS approximation, it would be possible to set the energy of the central emitter ``manually'' to that of the post-reaction HBQ state (i.e., resonant with the peripheral TLSs and/or the dipolar mode), this would overlook the substantial role played by the pseudo-mode in this configuration, as shown next.

We performed simulations with an initial excitation localized on solvated HBQ (cyclohexane, 300 K) at the center of the nanoparticle dimer gap (\figref{fig:hbq-meth-gap}{a}).
To provide additional context, in \figref{fig:hbq-meth-gap}{b} we show the \emph{effective} spectral density $J_{\text{HBQ}}(\omega)$ perceived by HBQ when treating the four MeB molecules as part of the ``photonic'' environment felt by the HBQ molecule within linear response.
Here, the four solvated Methylene Blue (MeB) molecules (water, 300 K) had distinct initial structures (static diagonal/off-diagonal disorder) and velocities (dynamic disorder), sampled from a Maxwell-Boltzmann distribution at 300 K.
Following blue/UV excitation of HBQ, its excited-state population (green) decreases monotonically (\figref{fig:hbq-meth-gap}{c}).
Concomitantly, the pseudo-mode population (black), which aggregates higher-order cavity modes~\cite{Delga2014} and several fitting modes in our few-mode approach (black dots in \figref{fig:hbq-meth-gap}{b}), increases, as does the ground-state occupation that reflects total losses.
Around $\sim$20 fs, the pseudo-mode population drops sharply while the bright dipolar mode (purple) rises by nearly an order of magnitude within $\sim$5 fs.
This transition coincides with completion of HBQ's intramolecular proton transfer (\figref{fig:hbq-meth-gap}{d}), after which the HBQ excited state experiences a Stokes shift into resonance with the dipolar mode.
Because MeB absorption overlaps with this dipolar mode (Fig. S9), the MeB ensemble becomes strongly coupled to it.
The MeB populations (light blue, orange, dark blue, and red) then begin coherent oscillations with the dipolar mode and rise by nearly an order of magnitude within another $\sim$5 fs.
By 100 fs, MeB populations reach $\sim$13\% of the total excitation, with MeB4 (red) reaching up to $\sim$8\% (\figref{fig:hbq-meth-gap}{c}).

When considering an idealized scenario in which the MeB emitters share identical initial structures at their absorption maximum (Fig. S10), the cavity-mediated energy transfer is not necessarily enhanced relative to the disordered case of \autoref{fig:hbq-meth-gap}.
This underscores the decisive role of dynamic disorder in ensembles of molecular emitters.

To further explore phase space for the HBQ + 4 MeB system, we performed additional simulations with different initial velocities sampled from a Maxwell-Boltzmann distribution at 300 K.
While these results (Fig. S11) corroborate the mechanism of cavity-mediated energy transfer between reactive HBQ and MeB, the total population accruing on the peripheral emitters remains small relative to overall losses (ground state reaching $\sim$70\%; yellow curve in \figref{fig:hbq-meth-gap}{c} and Fig, S11(h)).
 Future work will optimize this photochemistry-driven, cavity-mediated energy transfer setup \emph{in silico}.

\section{Conclusions}
Bridging recent advances in quantum optics~\cite{Medina2021,SanchezBarquilla2022} and quantum chemistry~\cite{Tichauer2021} into a unified numerical framework opens new opportunities for \emph{in silico} control of light-matter interactions at the nanoscale.
Leveraging high-performance computing and few-mode quantization of the electromagnetic field~\cite{Medina2021,SanchezBarquilla2022}, our implementation brings quantum optics and quantum chemistry together without sacrificing accuracy for multichromophoric systems coupled to confined fields in plasmonic nanostructures.
Using this tool, we show that strong coupling persists in the presence of substantial static and dynamic disorder introduced by nuclear motion, and that molecular degrees of freedom lift degeneracies preserved in TLS descriptions.
We further demonstrate cavity-mediated intermolecular energy transfer and highlight the role of pseudomodes in loss channels.
These advances pave the way for \emph{in silico} design of molecule-nanocavity architectures and the integration of organic emitters into photonic circuits~\cite{Toninelli2021}.

\section*{Acknowledgments}
This project has received funding from the European Union’s Horizon 2020 research and innovation programme under grant agreement No.\ 101034324 through a CIVIS3i postdoctoral fellowship to R.H.T.\ and under grant agreement No.\ 101070700 (MIRAQLS).
We also acknowledge financial support from the Spanish Ministry of Science, Innovation and Universities - Agencia Estatal de Investigación through the FPI Grant PRE2021-098978 to M.L.\ with support from ESF$+$, as well as Grants PID2021-125894NB-I00, EUR2023-143478, PID2024-161142NB-I00, and CEX2023-001316-M (through the María de Maeztu program for Units of Excellence in R\&D).

\printbibliography{}
\end{document}


\maketitle
\tableofcontents
\newpage

\section{Ehrenfest force in the adiabatic basis}
\label{si-sec:model}

In the theory section of the main text (section II), we presented the Hellmann-Feynman force for mean-field Ehrenfest dynamics~\cite{Ehrenfest1927} in the diabatic basis of uncoupled molecular and photonic excitations.
We here additionally show the same forces when working in the \emph{adiabatic} basis of the light-matter Hamiltonian consisting of its (polaritonic) eigenstates, $|\Psi(t)\rangle = \sum_{p=1}^{N+n} c_p(t) |\phi_p\rangle$, with $|\phi_p\rangle = \sum_{m=1}^N \beta_m^p |\psi_m\rangle + \sum_{m=N+1}^{N+n} \alpha_k^p |\psi_m\rangle$. The forces are then given by
\begin{equation}
    \begin{split}
        \mathbf{F}^{\mathrm{adiabatic}}_{{HF}\to a\in m}=&\left[|c_p(t)|^2|\beta_m^p|^2+\left(c_p^*(t)c_q(t)\beta_m^{p*}\beta_m^q+c.c.\right)\right]\nabla_{a\in m}V_{{S}_1}({\bf{R}}_m) \\
        +& \left[ |c_p(t)|^2\left( \sum_{j\neq m} |\beta_j^p|^2+\sum_k|\alpha_k^p|^2\right)-\left(c_p^*(t)c_q(t)\beta_m^{p*}\beta_m^q+c.c.\right)\right]\nabla_{a\in m}V_{{S}_0}({\bf{R}}_m) \\
        -& |c_p(t)|^2\left( \beta_m^{p*}\sum_k\alpha_k^p + c.c. \right)\nabla\mathbf{\mu}({\bf{R}}_m)\cdot \mathbf{g}_{jk}\\
        -& c_p^*(t)c_q(t) \sum_k \left(\beta_m^{p*} \alpha_k^q +\beta_m^q \alpha_k^{p*} \right) \nabla\mathbf{\mu}({\bf{R}}_m)\cdot \mathbf{g}_{jk} \\
        -& c_p(t)c_q^*(t) \sum_k \left(\beta_m^p \alpha_k^{q*} + \beta_m^{q*}\alpha_k^p \right) \nabla\mathbf{\mu}({\bf{R}}_m)\cdot \mathbf{g}_{jk}.
    \end{split}
\end{equation}
    
\section{Validation of the few-mode QM/MM MD implementation}
\subsection{Comparison with the Lindblad master equation}
\begin{figure}[htb!]
\includegraphics[width=\textwidth]{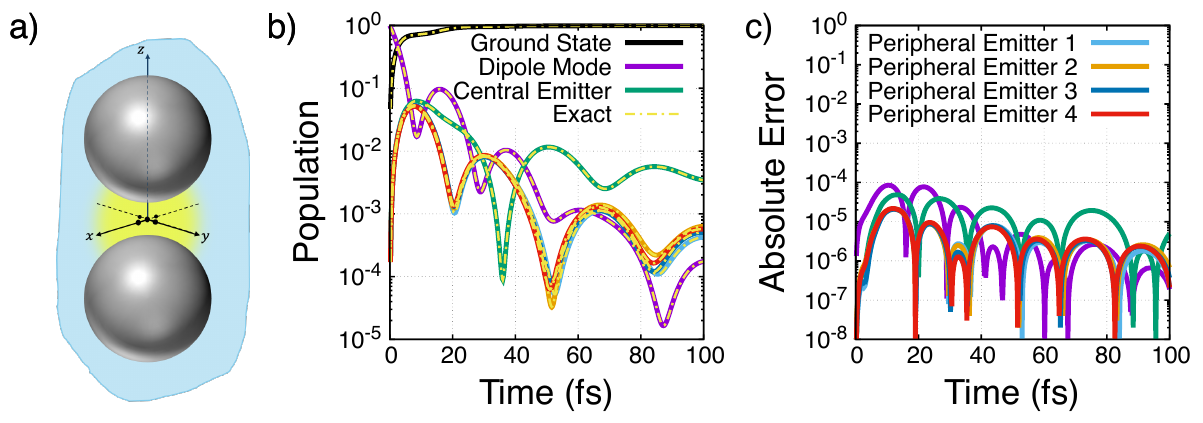}
\caption{(a) Schematic representation of five point-like emitters, modeled as ideal two-level systems, within the gap of a silver nanoparticle dimer oriented along the $z$-axis.
(b) Time evolution of the populations of the five emitters resonantly coupled to the dipolar mode of the optical resonator. 
(c) Absolute error between the population dynamics obtained with our integrator and with the Lindblad master equation, labeled as the ``exact'' reference method.}
\label{si-fig:errortls}
\end{figure}
To validate the present implementation that incorporates QM/MM MD into the few-mode approach~\cite{Medina2021,SanchezBarquilla2022}, we start by modeling molecular emitters as ideal TLSs within the setup depicted in \figref{si-fig:errortls}{a}. 
By integrating the time-dependent Schrödinger equation, in either the diabatic or locally diabatic basis~\cite{Granucci2001} (the latter is employed when working in the adiabatic basis of $H^{LM}$, Eq. (6)), we compute the time evolution of the total (polaritonic) wavefunction of the hybrid light-matter system
\begin{equation}
|\Psi(t)\rangle = \sum_{p=1}^{N+n}c_p(t) |\psi^p\rangle
\end{equation} 
In the above expression, $c_p(t)$ are the time-dependent expansion coefficients and $\psi^p$ are the basis states. 

We then compute the populations $\rho_p(t)=c_p^*(t)c_p(t)$ of the diabatic or adiabatic states $\psi^p$. 
Comparison between the population dynamics obtained with the present implementation and the well-established Lindblad master equation of open quantum systems for TLSs~\cite{Breuer2007} shows that the absolute difference between these two integrators is on the order of $10^{-4}$-$10^{-5}$, as depicted in \figref{si-fig:errortls}{b} and \figref{si-fig:errortls}{c}.

\subsection{Energy conservation}
\begin{figure}[htb!]
\includegraphics[width=\textwidth]{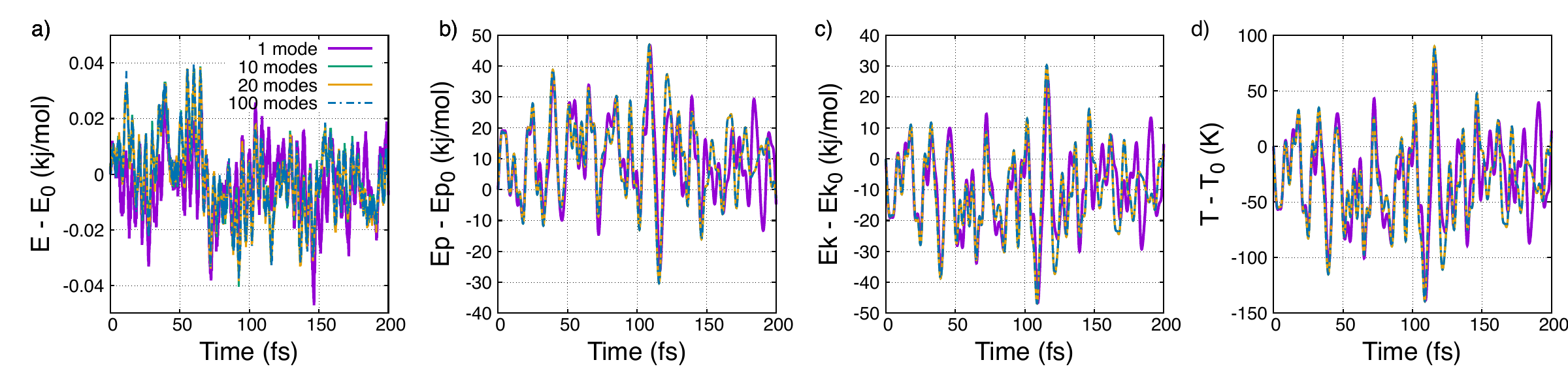}
\caption{Fluctuations in the (a) total, (b) potential, and (c) kinetic energy, as well as in the (d) temperature, during few-mode (full-QM) MD simulations of one Rhodamine chromophore in gas phase coupled to lossless 2D Fabry-Pérot microcavities described by 1 (purple), 10 (green), 20 (orange), and 100 (blue dashed) cavity modes. 
These simulations were run in the diabatic basis of uncoupled molecular and photonic excitations ($|S_0^0 \ldots S_1^i \ldots S_0^N\rangle \otimes |0_0 \ldots 1_k \ldots 0_n\rangle$).}
\label{fig:nmodes}
\end{figure}

\begin{figure}[htb!]
\includegraphics[width=\textwidth]{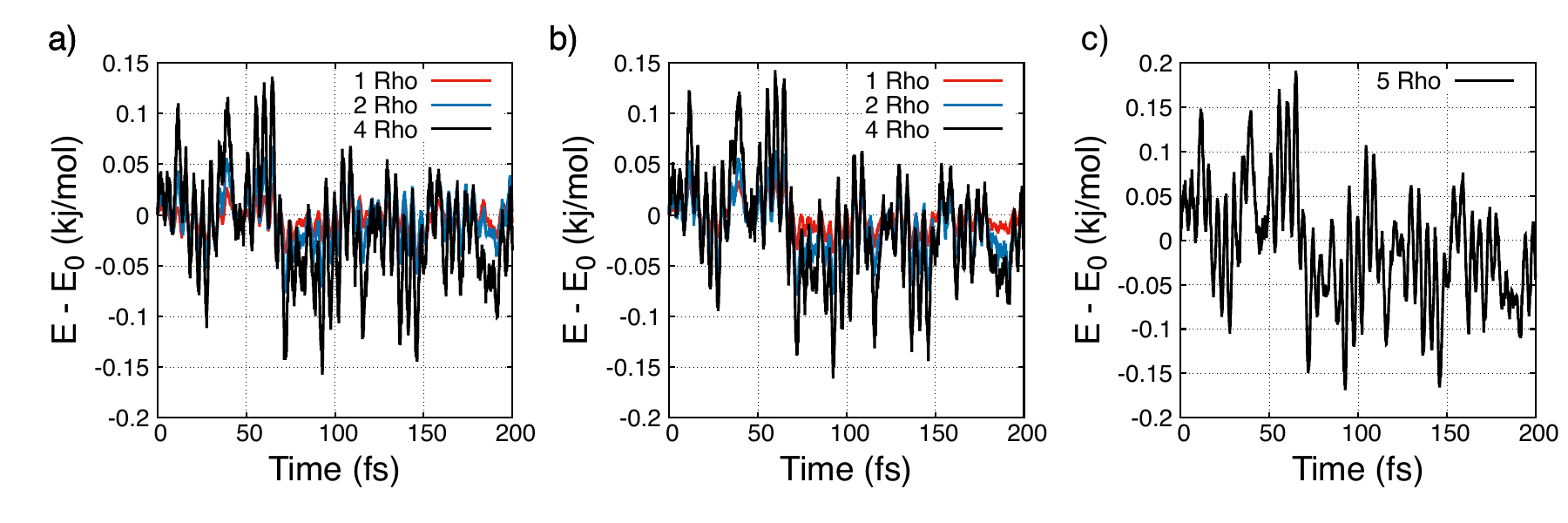}
\caption{Fluctuations in the total energy during few-mode (full-QM) MD simulations of one (red), two (blue), and four (black) Rhodamine chromophores in gas phase coupled to lossless 2D Fabry-Pérot microcavities described by (a) a single cavity mode and (b) 10 cavity modes. The simulations in (a) were run in the diabatic basis of uncoupled molecular and photonic excitations ($|S_0^0 \ldots S_1^i \ldots S_0^N\rangle \otimes |0_0 \ldots 1_k \ldots 0_n\rangle$) while the simulations in (b) were run in the adiabatic basis of the light-matter Hamiltonian, Eq. (6). (c) Fluctuations of the total energy of the system  depicted in \autoref{fig:enerdiff}{a} when omitting losses of the nanoresonator.}
\label{fig:enertot}
\end{figure}

As previously done and for comparison with the former implementation beyond a single-mode approximation~\cite{Tichauer2021}, we test the numerical implementation of the forces by performing full-QM gas-phase Ehrenfest simulations (time step $dt=0.1$ fs) of one, two, and four Rhodamine chromophores ($S_0$ and $S_1$ states described at the HF/3-21G and CIS/3-21G levels of theory, respectively) coupled to lossless 2D Fabry-Pérot microcavities described by 1, 10, 20, and 100 modes ($\omega_0=3.95$ eV, $L=5$ $\mu$m length, field amplitude $E_0=0.002$ au).  In \autoref{fig:nmodes} and \autoref{fig:enertot} we show that the total energy is conserved, i.e., no energy drift as a function of time. Importantly, the energy fluctuations are independent of the number of confined photon modes, i.e., same amplitude for cavities described by 1, 10, 20, and 100 cavity modes, as depicted in \autoref{fig:nmodes}.

Running simulations in the adiabatic basis of the light-matter Hamiltonian, Eq. (6), is more computationally demanding than running the same simulation in the diabatic basis of uncoupled molecular and photonic excitations ($|S_0^0 \ldots S_1^i \ldots S_0^N\rangle \otimes |0_0 \ldots 1_k \ldots 0_n\rangle$~\cite{Tichauer2021,Sokolovskii2024}) because of the step involved in diagonalizing the matrix. 
We ensure that energy is conserved irrespective of the chosen basis. 
In \autoref{fig:enertot}{b} we present energy conservation tests for simulations in the adiabatic basis.

\begin{figure}[htb!]
\includegraphics[width=\textwidth]{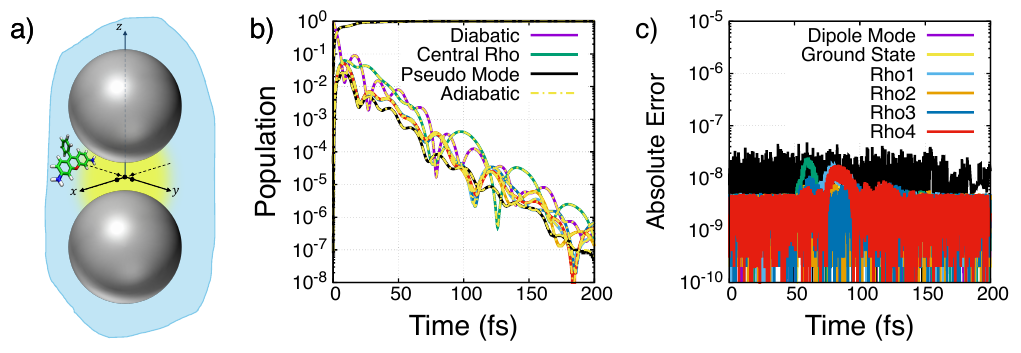}
\caption{(a) Schematic representation of five Rhodamine chromophores within the gap of a silver nanoparticle dimer oriented along the $z$-axis. 
(b) Excited-state populations of the central (green) Rhodamine, the peripheral (light blue, orange, blue, and red) molecular emitters, and the dipolar mode of the nanoresonator (purple) computed from simulations in the diabatic basis of uncoupled molecular and photonic excitations ($|S_0^0 \ldots S_1^i \ldots S_0^N\rangle \otimes |0_0 \ldots 1_k \ldots 0_n\rangle$). 
These same populations obtained from simulations in the adiabatic basis of the light-matter Hamiltonian, Eq. (6), are plotted with yellow dashed lines. 
(c) Absolute difference in the populations when running simulations in the diabatic and adiabatic basis.}
\label{fig:enerdiff}
\end{figure}

We also performed gas-phase (full QM) MD simulations at 0 K of five Rhodamine chromophores coupled to the field confined by a silver nanoparticle dimer. 
As depicted in \autoref{fig:enerdiff}, the absolute difference in the excited-state populations are on the order of $10^{-8}$-$10^{-7}$ when running in the diabatic and adiabatic basis of the light-matter Hamiltonian (Eq. (6)). When omitting the losses of the nanoresonator, the total energy is conserved (\autoref{fig:enertot}{c})

\section{Additional results}
\subsection{Effect of incorporating molecular degrees of freedom}
\begin{wrapfigure}[14]{r}{0.35\textwidth}
\vspace{-4mm}
\includegraphics[width=\textwidth]{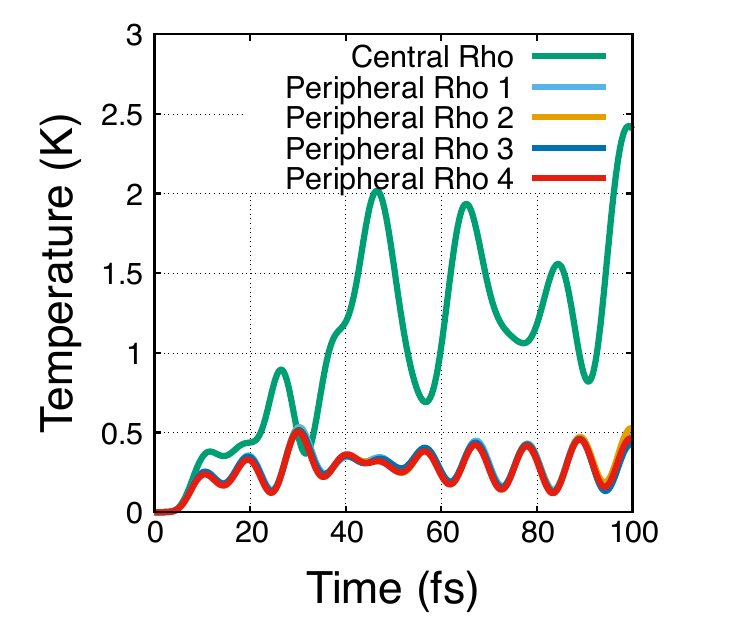}
\caption{\small Temperature during gas-phase few-mode (full QM) MD simulations at 0 K.}
\label{si-fig:temp}
\end{wrapfigure}
We performed numerical simulations of the hybrid light-matter system depicted in Fig. 1a of the main article with identical initial conditions (excitation energies, transition dipole moments) but treating the five Rhodamine molecules as \emph{(i)} ideal two-level systems (Fig. 1 in the main), \emph{(ii)} in the gas phase (full QM) at different temperatures (Fig. 2 in the main), and \emph{(iii)} in solution (QM/MM) at 300 K (Fig. 3 in the main). 

For controlling temperature during MD simulations, we employed the velocity rescaling algorithm~\cite{Busi2007} with a time constant coupling of 0.1 ps applied every 100 steps of the simulation (time step $dt=0.1$ fs).
During the few-mode (full QM) gas-phase simulation at 0 K, the temperature of the different emitters is slightly higher than that set by the MD thermostat but stabilizes around a fixed value: approximately 0.25 K for the peripheral and approximately 1.5 K for the central Rhodamine as depicted in \autoref{si-fig:temp}.  

\subsection{Initially exciting the central emitter}
\begin{figure}[htb!]
\includegraphics[width=\textwidth]{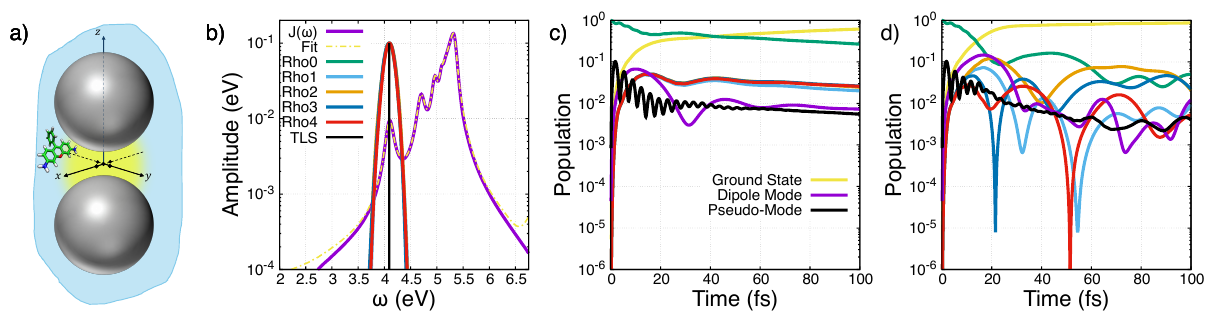}
\caption{(a) Schematic representation of five point-like emitters within the gap of a silver nanoparticle dimer oriented along the $z$-axis. 
(b) Physical (purple) and fitted (yellow dashed) spectral densities of the central emitter with its dipole moment oriented along the $z$-axis plotted together with the absorption spectra of the Rhodamine model (green, light blue, orange, dark blue, and red). 
When treated as ideal two-level systems (TLS), the absorption of these molecular emitters corresponds to a delta function (black vertical line). 
(c) Time evolution of the populations of the five emitters, modeled as ideal TLS, resonantly coupled to the dipolar mode of the optical resonator when initially exciting the central emitter. 
(d) Same populations but accounting for the molecular degrees of freedom during a few-mode QM/MM MD simulation with identical initial parameters as the TLS treatment.}
\label{si-fig:gapRho0}
\end{figure}

Following Gonzalez-Ballestero et al.~\cite{GonzalezBallestero2015}, we investigate cavity-mediated intermolecular energy transfer within a two-level system approximation (\figref{si-fig:gapRho0}{c}) and including the molecular degrees of freedom (Fig. 4 in the main article and \figref{si-fig:gapRho0}{d}). 

While the simulation presented in \autoref{si-fig:gapRho0}(d) does not correspond to the best case scenario where dynamic disorder enhances intermolecular energy transfer maximally, the QM/MM MD treatment of the emitters leads to higher occupation of one of the peripheral emitters (approximately 10\% at 20 fs) than the two-level system treatment (approximately 4\% at 16-17 fs).
This is due to the broadening of the molecular resonance that thus has higher overlap with the dipole mode of the nanoresonator, resulting in stronger light-matter coupling.

\subsection{Photochemistry-driven energy transfer}
\begin{wrapfigure}[12]{r}{0.35\textwidth}
\vspace{-16mm}
\includegraphics[width=0.95\textwidth]{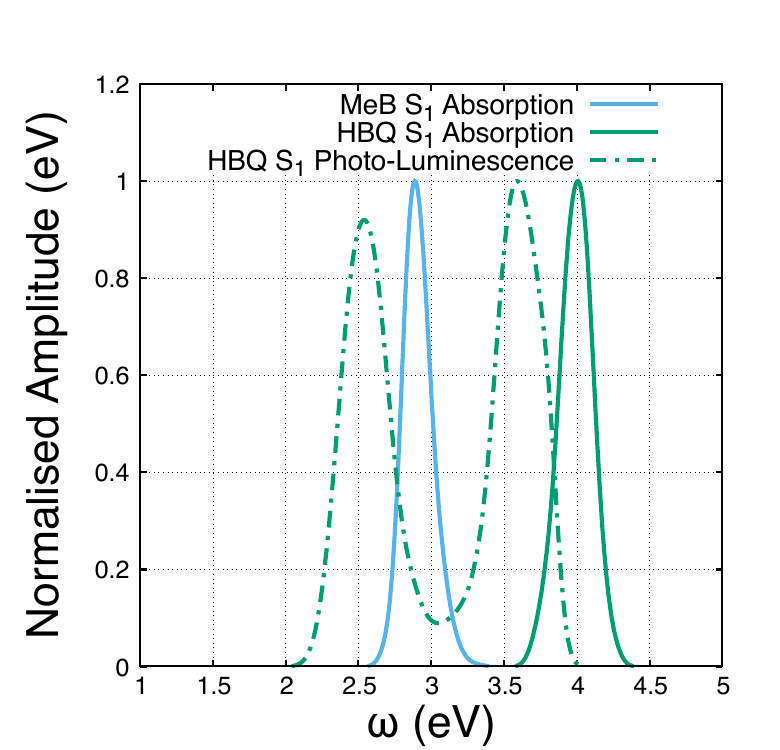}
\vspace{-1mm}
\caption{\small Normalized absorption and photoluminescence spectra of Methylene Blue and HBQ.}
\label{si-fig:hbqmeb}
\end{wrapfigure}
As shown in \autoref{si-fig:hbqmeb}, the overlap between the QM/MM absorption spectrum of Methylene Blue ($\omega$B97X-D/6-31G*/TIP3P TD-DFT level of theory) and the photoluminescence (PL) spectrum of HBQ (CAM-B3LYP/6-31G*/Gromos-2016H66 TD-DFT level of theory) makes these two molecular emitters suitable models for investigating, \emph{in silico}, photochemistry-driven donor-acceptor intermolecular energy transfer mediated by cavity modes. 

We note that the photoluminescence spectrum of HBQ strongly depends on when the intramolecular proton transfer reaction takes place.
The sooner this chemical reaction completes, the higher the left peak and the smaller the right peak in its PL spectrum.
\begin{figure}[htb!]
\includegraphics[width=\textwidth]{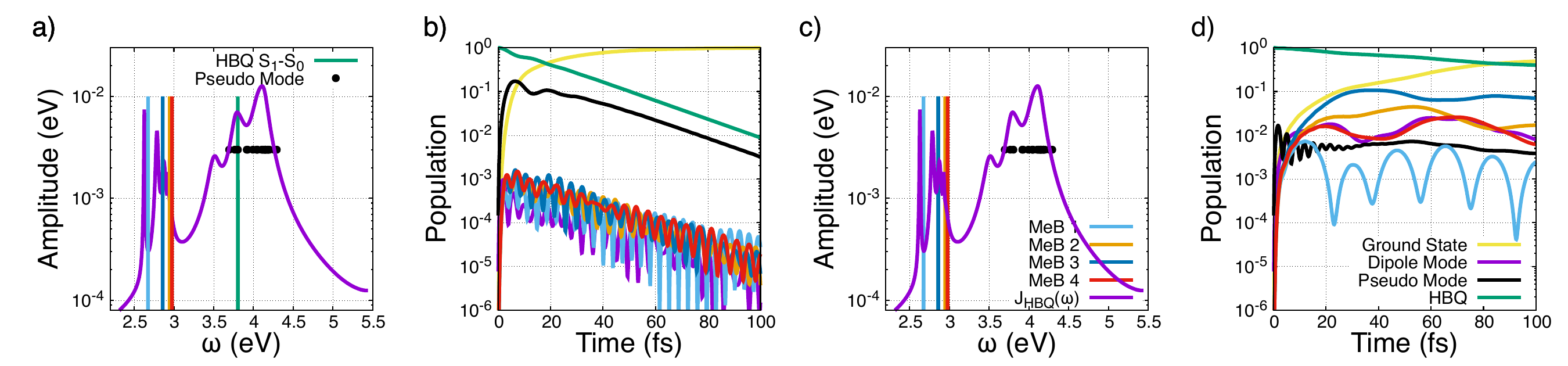}
\caption{(a) Effective spectral density, perceived by HBQ, plotted together with the absorption spectra of Methylene Blue (light blue, orange, dark blue, and red). 
Black dots indicate the frequencies $\tilde{\omega}_k$ of the fitted modes that form the pseudo-mode in our model. 
(b) Time evolution of the excited-state population of the five emitters (HBQ green, MeB1 light blue, MeB2 orange, MeB3 dark blue, and MeB4 red) together with that of the dipolar mode (purple) and the pseudo-mode (black) of the nanoresonator. 
The ground-state population (yellow) increases steadily due to losses in the system. 
(c) and (d) same as panels (a) and (b), respectively, but after HBQ has undergone the intramolecular proton transfer reaction.}
\label{si-fig:hbqmebTLS}
\end{figure}

We performed simulations for investigating photochemistry-driven energy transfer within a two-level system treatment of the molecular emitters.
As shown in \figref{si-fig:hbqmebTLS}{b}, such a treatment precludes energy transfer because of the lack of overlap in the initial excitation energies of both emitters (see spectrum in \figref{si-fig:hbqmebTLS}{a}).
Running such TLS simulations with the central TLS corresponding to HBQ after the excited-state intramolecular proton transfer, i.e., central TLS resonant with the four peripheral ones, dismisses the large impact of the so-called pseudo-mode~\cite{Delga2014} in this setup (\figref{si-fig:hbqmebTLS}{d}) as we show in Fig. 5 of the main article.

\begin{figure}[htb!]
\floatbox[{\capbeside\thisfloatsetup{capbesideposition={right,top},capbesidewidth=5cm}}]{figure}[\FBwidth]
{\caption{Spectral density of the silver nanoparticle dimer with five emitters in the center of its gap together with the normalized absorption and photoluminescence spectra of Methylene Blue and HBQ with (a) and without (b) initial disorder in the Methylene Blue ensemble.}\label{si-fig:spec-hbq-meb-gap}}
{\includegraphics[width=0.68\textwidth]{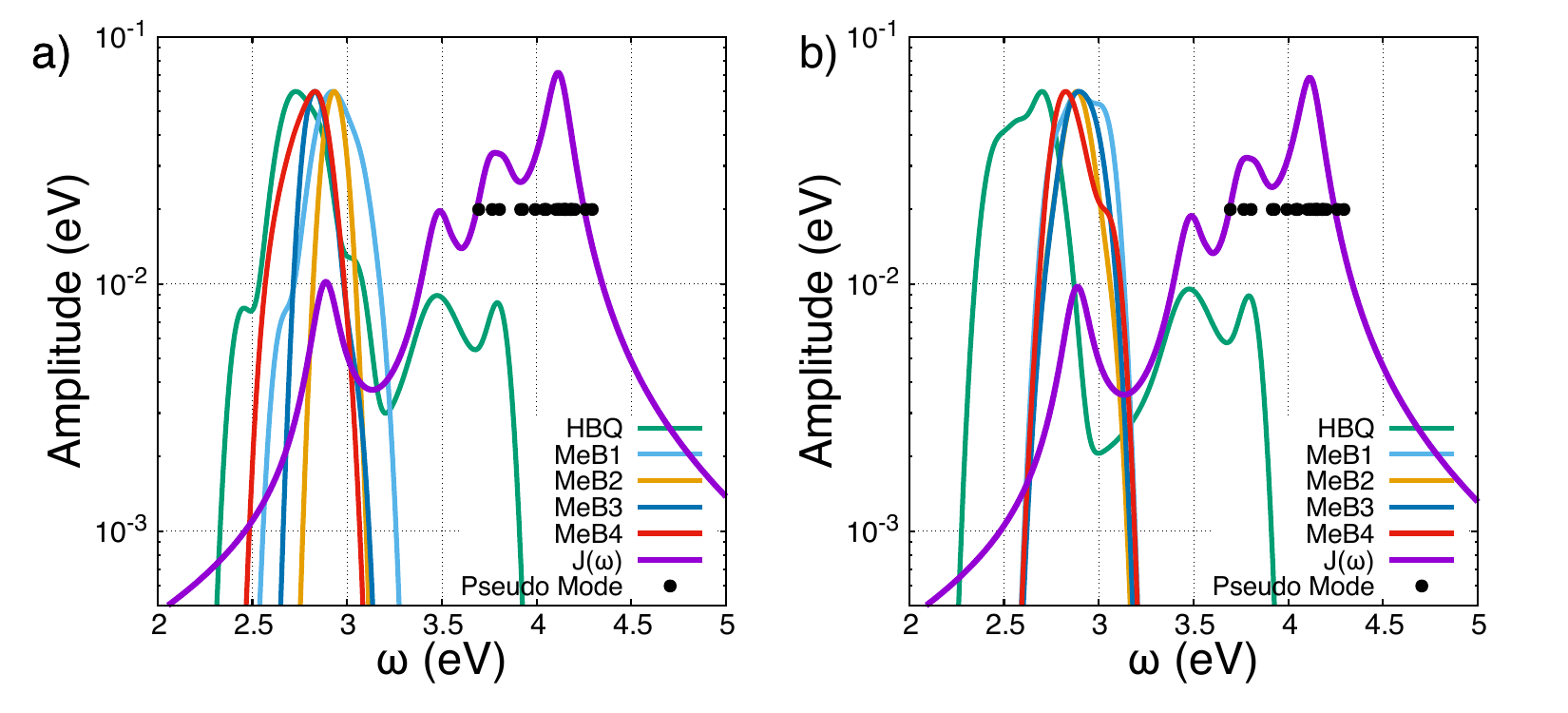}}
\end{figure}
For the MD simulation presented in the main article, dynamic disorder leads to a convenient overlap between the dipole mode of the silver nanoparticle dimer, the absorption of Methylene Blue, and the emission of HBQ as depicted in \figref{si-fig:spec-hbq-meb-gap}{a}, in contrast to the same simulation but without initial diagonal/off-diagonal disorder in the Methylene Blue ensemble, \figref{si-fig:spec-hbq-meb-gap}{b}.
Despite identical initial conditions for HBQ (same atomic positions and velocities in both simulations), HBQ Stokes shifts more significantly after the proton transfer reaction in the simulation without initial disorder in the Methylene Blue ensemble than in the simulation with initial disorder.
This leads to reduced overlap of HBQ emission with the dipole mode of the nanoresonator and the absorption of Methylene Blue.

\begin{figure}[htb!]
\includegraphics[width=\textwidth]{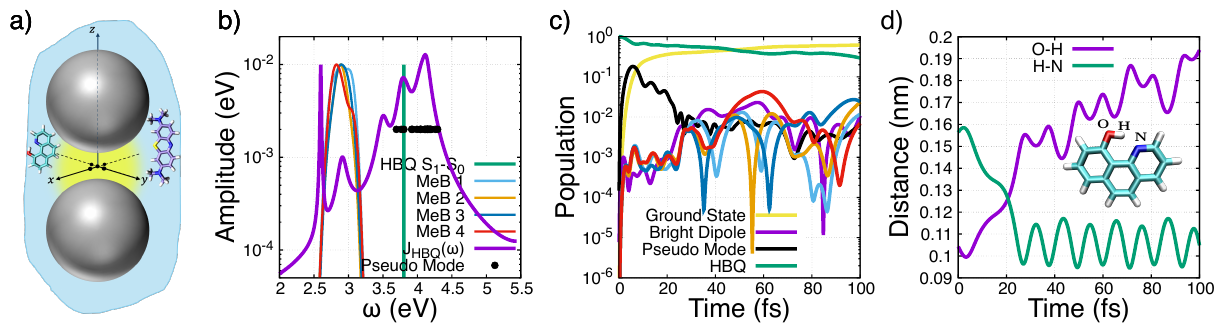}
\caption{(a) Schematic representation of five molecular emitters within the gap of a silver nanoparticle dimer oriented along the $z$-axis. 
A molecular model of HBQ, depicted in the left inset, is positioned at the central spot while Methylene Blue molecules, depicted in the right inset, occupy the peripheral positions. 
(b) Effective spectral density, perceived by HBQ, plotted together with the absorption spectra of Methylene Blue (light blue, orange, dark blue, and red). 
Black dots indicate the frequencies $\tilde{\omega}_k$ of the fitted modes that form the pseudo-mode in our model. 
(c) Time evolution of the excited-state population of the five emitters (HBQ green, MeB1 light blue, MeB2 orange, MeB3 dark blue, and MeB4 red) together with that of the dipolar mode (purple) and the pseudo-mode (black) of the nanoresonator. 
The ground-state occupation (yellow) increases steadily due to losses in the system. 
(d) Distances between O, H, and N atoms of HBQ (inset). 
A distance of approximately 0.105 nm between H and N corresponds to a bonded state of the two atoms.}
\label{si-fig:idealhbqmeb}
\end{figure}
Indeed, when running a few-mode QM/MM MD simulation with idealized starting conditions where all Methylene Blue molecules have the same initial conformation and thus no initial diagonal/off-diagonal disorder (\autoref{si-fig:idealhbqmeb}), the cavity-mediated intermolecular energy transfer between the initially excited HBQ and the four peripheral Methylene Blue molecules is not necessarily enhanced compared to a situation with initial disorder as shown in Fig. 5 of the main article.
This highlights the paramount effect of dynamic disorder and consequently the need for dynamic approaches with which multiple trajectories can be run to sample the enormous phase space available to the system.

\begin{figure}[htb!]
\includegraphics[width=\textwidth]{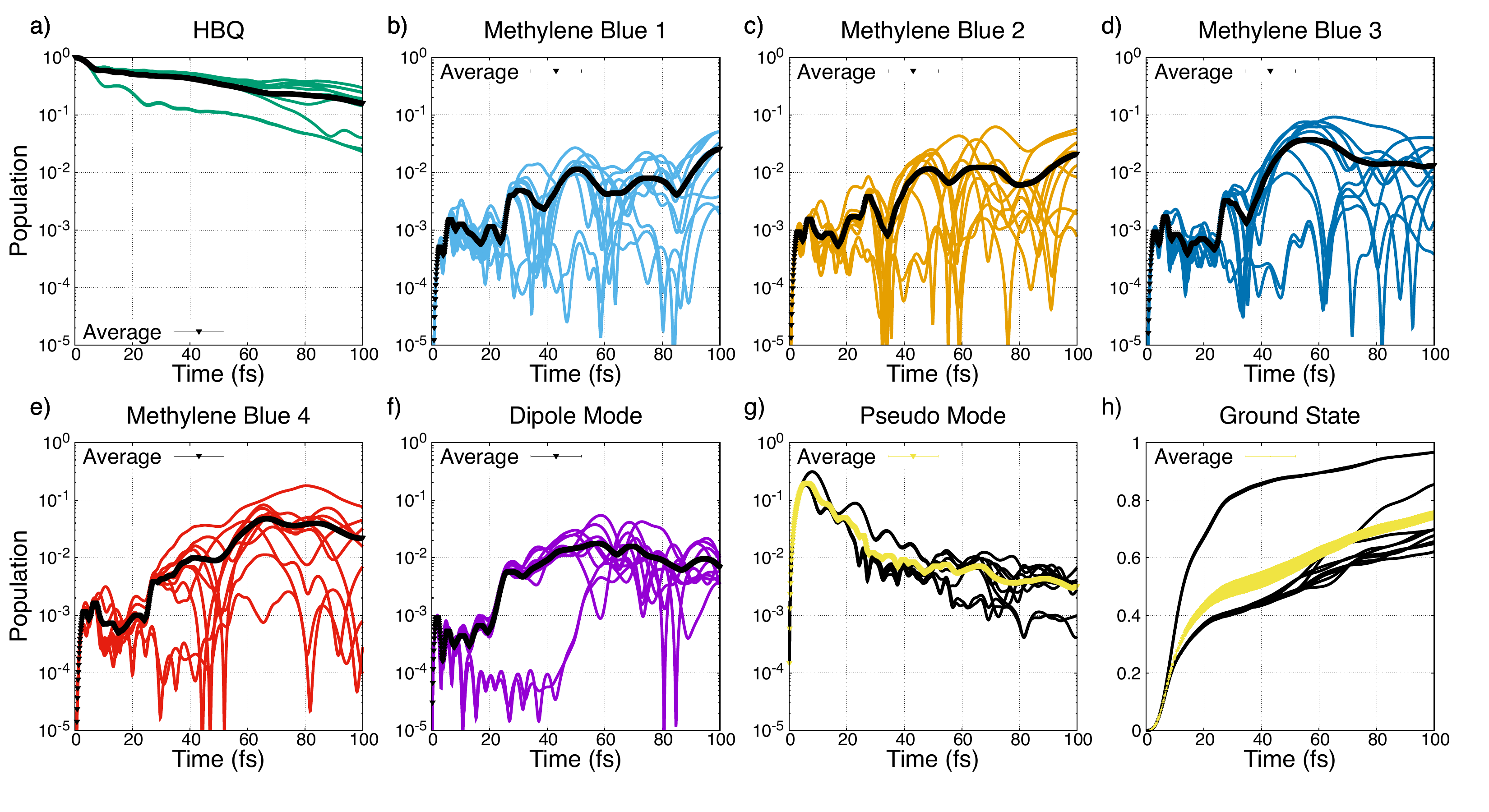}
\caption{Time evolution of the excited-state population during 10 few-mode QM/MM MD simulations of (a) an HBQ molecule positioned at the central spot within the setup depicted in \autoref{si-fig:idealhbqmeb} and of (b), (c), (d), (e) Methylene Blue molecules at the peripheral positions. 
The average over these 10 realizations is plotted with black diamonds. 
Population dynamics of (f) the dipolar mode and (g) the modes constituting the broad pseudo-mode of the silver nanoparticle dimer. 
(h) Ground-state occupation due to losses of the silver nanoparticle dimer. 
Note the linear scale in this last plot. 
The average together with the standard deviation is plotted with yellow diamonds.}
\label{si-fig:manyhbqmeb}
\end{figure}
We conducted additional few-mode QM/MM MD simulations presented in \autoref{si-fig:manyhbqmeb}.
These additional results confirm the underlying mechanism involved in photochemistry-driven, cavity-mediated intermolecular energy transfer, with the efficiency depending on \emph{(i)} the time at which the intramolecular proton transfer of HBQ completes and \emph{(ii)} the overlap between the emission of HBQ, the resonant/off-resonant coupling to the dipolar mode of the nanoresonator, and the absorption of Methylene Blue.
Importantly, this last factor is highly influenced by dynamic disorder.

\newpage
\printbibliography{}